# A Large Iron Isotope Effect in $SmFeAsO_{1-x}F_x$ and $Ba_{1-x}K_xFe_2As_2$

R. H. Liu[1], T. Wu[1], G. Wu[1], H. Chen[1], X. F. Wang[1], Y. L. Xie[1], J. J. Yin[1], Y. J. Yan[1], Q. J. Li[1], B. C. Shi[1], W. S. Chu[2], Z. Y. Wu[2], X. H. Chen[1]

*1.* Hefei National Laboratory for Physical Sciences at Microscale and Department of Physics, University of Science and Technology of China, Hefei, Anhui 230026, China

*2.* Beijing Synchrotron Radiation Facility, Institute of High Energy Physics, Chinese Academy of Sciences, Beijing 100049, China and National Synchrotron Radiation Laboratory, University of Science and Technology of China, Hefei 230026, China

**The recent discovery of superconductivity in oxypnictides with the critical temperature ($T_C$) higher than McMillan limit of 39 K (the theoretical maximum predicted by Bardeen-Cooper-Schrieffer (BCS) theory) has generated great excitement[1-5]. Theoretical calculations indicate that the electron-phonon interaction is not strong enough to give rise to such high transition temperatures[6], while strong ferromagnetic/antiferromagnetic fluctuations have been proposed to be responsible[7-9]. However, superconductivity and magnetism in pnictide superconductors show a strong sensitivity to the lattice, suggesting a possibility of unconventional electron-phonon coupling. Here we report the effect of oxygen and iron isotopic mass on $T_C$ and the spin-density wave (SDW) transition temperature ($T_{SDW}$) in $SmFeAsO_{1-x}F_x$ and $Ba_{1-x}K_xFe_2As_2$ systems. The results show that oxygen isotope effect on $T_C$ and $T_{SDW}$ is very little, while the iron isotope exponent $\alpha_C$=-d$lnT_C$/d$lnM$ is about 0.35, being comparable to 0.5 for the full isotope effect. Surprisingly, the iron isotope exchange shows the same effect on $T_{SDW}$ as $T_C$. These results indicate that electron-phonon interaction**

**plays some role in the superconducting mechanism, but simple electron-phonon coupling mechanism seems to be rather unlikely because a strong magnon-phonon coupling is included. Sorting out the interplay between the lattice and magnetic degrees of freedom is a key challenge for understanding the mechanism of high-$T_C$ superconductivity.**

Recent inelastic neutron scattering measurements on $Ba_{1-x}K_xFe_2As_2$ (x=0 and 0.4) provide evidence for presence of magnetic excitations[10,11]. It suggests that spin fluctuation may play an important role for the mechanism of superconductivity. However, it is demonstrated that phonons couple selectively to the spin system[12]. The structural transition from tetragonal to orthorhombic is driven by the antiferromagnetic SDW order[13], and the antiferromagnetic SDW exists only in the orthorhombic structure[14,15]. The pressure coefficient of $T_C$, $dT_C/dP$, changes from positive to negative with a crossover from orthorhombic to tetragonal symmetry for the superconducting phase[14]. The superconductivity and spin-density wave coexist in the orthorhombic structure[14,16,17]. These results indicate remarkable sensitivity of superconductivity and magnetism to the lattice.

The isotopically substituted polycrystalline samples with nominal compositions $SmFeAsO_{1-x}F_x$ (x=0, 0.15) and $Ba_{1-x}K_xFe_2As_2$ (x=0, 0.4) were synthesized by conventional solid state reaction described in Ref.5 and Ref.17, respectively. Figure 1 shows the Raman spectra for the samples $SmFeAsO_{1-x}F_x$ by replacing $^{16}O$ with $^{18}O$, and for the samples $SmFeAsO_{1-x}F_x$ and $Ba_{1-x}K_xFe_2As_2$ by replacing $^{56}Fe$ with the isotope $^{54}Fe$. The frequency shift of 4.2% and 4.5% for the $E_g$ mode of oxygen[18] around 420 cm$^{-1}$ suggests about 71% and 77% $^{18}O$ substitution for $^{16}O$ for x=0 and 0.15 samples, respectively. Raman shift of about 1.7% observed in the four samples for the $B_{1g}$ mode of iron[18] indicates almost 100% $^{54}$Fe substitution for $^{56}$Fe for the two systems. These data are listed in Table 1.

The temperature dependence of resistivity (ρ) and its derivative (dρ/dT) for typical samples SmFeAsO$_{1-x}$F$_x$ on replacing $^{16}$O with the isotope $^{18}$O are shown in Fig. 2. T$_C$ and T$_{SDW}$ are listed in Table 2 for all samples from different batches. Based on isotope exponent α$_C$=-d$lnT_C$/d$lnM$, the α$_C$ is deduced to be -0.06(1) for the superconducting transition. To quantitatively compare the isotope effect on T$_{SDW}$ with on T$_C$, we also define an isotope exponent α$_{SDW}$=-d$lnT_{SDW}$/d$lnM$ for SDW transition although no theory is established for isotope effect on magnetic phase transition yet. α$_{SDW}$=-0.05(1) is obtained. These results indicate that oxygen isotope effect on T$_C$ and T$_{SDW}$ is very little.

Temperature dependence of resistivity and its derivative for typical samples SmFeAsO$_{1-x}$F$_x$ and Ba$_{1-x}$K$_x$Fe$_2$As$_2$ by substitution of $^{54}$Fe for $^{56}$Fe are shown in Fig.3. An increase in T$_C$ is clearly observed in resistivity measurements, and dρ/dT clearly shows an increase of SDW transition by substitution of $^{54}$Fe for $^{56}$Fe. The average results for several different samples are listed in Table 2. The average isotope component α$_{SDW}$ for several samples of SmFeAsO and BaFe$_2$As$_2$ is 0.39(2) and 0.36(2), and α$_C$ is found to be 0.34(3) and 0.37(3) averaged over several samples of SmFeAsO$_{0.85}$F$_{0.15}$ and Ba$_{0.6}$K$_{0.4}$Fe$_2$As$_2$, being comparable to 0.5 for full isotope effect in the framework of BCS theory. It indicates a strong iron isotope effect on T$_C$ and T$_{SDW}$. It implies that electron-phonon interaction should play an important role for the superconducting mechanism. It is striking that the iron isotope exponents (α$_C$ and α$_{SDW}$) on Tc and T$_{SDW}$ are almost the same for the two systems, and much larger than the oxygen isotope exponents.

Isotope effect studies require well-characterized samples with reproducible crystal chemistry properties. An important experimental point has to be addressed about the sample processing. It is found that $T_{SDW}$ is insensitive to the sample processing for the parent compounds. $T_C$ of the sample $SmFeAsO_{0.85}F_{0.15}$ is sensitive to the sample processing because the F content is not easy to control. A detailed description of the synthesis procedure used to ensure the same F content is given in the Supplementary Information. No difference in the lattice constants (see Supplementary Fig. S1) provides strong evidence for the same F content for isotope exchange. To confirm that the observed results is intrinsic instead of impurity effect, we checked the difference of $T_C$ and $T_{SDW}$ for the samples $Ba_{1-x}K_xFe_2As_2$ with natural abundance iron ( $^n$Fe ) with purity of 99.9% and $^{56}$Fe with purity of 99.78%. $T_{SDW}$ and $T_C$ are nearly the same for the samples with $^n$Fe and $^{56}$Fe (see Supplementary Fig. S3). We synthesized the samples $Ba_{0.6}K_{0.4}Ba_2As_2$ using $^n$Fe with purity of 98% and 99.9% to check impurity effect on Tc. The difference of Tc for the two samples is 0.07 K (see Supplementary Fig. S4). It indicates that the effect of impurity on Tc is very little, and does not affect the intrinsic isotope effect observed in Table 2.

It should be emphasized that iron isotope exchange has a strong effect on spin-density-wave state. Substitution of $^{54}$Fe for $^{56}$Fe leads to a remarkable decrease in resistivity below the SDW ordering temperatures with a large $\alpha_{SDW}$ for the two systems. It suggests a strong magnon-phonon coupling. A giant oxygen isotope effect has been observed in magnetoresistive $La_{1-x}Ca_xMnO_{3+y}$, and the isotope exponent $\alpha_{FM}$ for ferromagnetic transition is as high as 0.85[21]. Such a large isotope shift is believed to arise from coupling of the charge carriers to Jahn-Teller lattice distortions[21]. In pnictide superconductors, the strong sensitivity of superconductivity and magnetism to the lattice may be responsible for the large isotope effect. These results definitely indicate that electron-phonon interaction plays an important role for the superconducting mechanism, but the strong magnon-phonon coupling has to be considered.

The iron isotope effect on $T_{SDW}$ and $T_C$ is much larger than the oxygen isotope effect in pnictide superconductors. The reason could be that the iron-arsenide plane is conducting layer and responsible for the superconductivity, and SDW ordering originates from Fe moment. For $MgB_2$ superconductor, no magnetic correlation is included and the superconductivity can be understood with BCS theory with $\alpha_c=0.32$[22]. In cuprates, the isotope effect on $T_C$ is sensitive to doping level. The effect is vanishing at optimum doping, but increases systematically with decreasing doping level to be maximum at the border to the antiferromagnetic state[23,24]. It seems that the isotope effect is somewhat related to magnetic fluctuation. Such unconventional isotope effects demonstrate that the electron–phonon interaction also plays an important role in the physics of cuprates. Sorting out the interplay between the lattice and magnetic degrees of freedom is a key challenge for understanding the mechanism of high-$T_C$ superconductivity.

**Supplementary Information** is linked to the online version of the paper at www.nature.com/nature.

**Acknowledgments:** This work is supported by the Nature Science Foundation of China, and by the Ministry of Science and Technology of China and Chinese Academy of Sciences. We acknowledge Prof. Z. X. Shen for helpful discussion and encouragement, and Profs. D. L Feng and S. Y. Li for discussions.

**Author Contributions:** X.H.C. designed and coordinated the whole experiment, and analysed the data and wrote the paper. R.H.L. and T.W. did the main experiments including sample preparation and analysed the data. G.W., X.F.W. and B.C.S synthesized the samples. H.C. and Y.L.X. partially measured the resistivity. J.J.Y. measured the susceptibility. Y.J.Y. and Q.J.L. did X-ray powder diffraction measurements. W.S.C. and Z.Y.W. provided the iron isotope $^{54}$Fe.

**Author information:** The authors declare no competing financial interests. Correspondence and requests for materials should be addressed to X. H. Chen (chenxh@ustc.edu.cn).

Table 1: **Raman shifts of the $E_g$ mode of oxygen ($\omega_O$) and the $B_{1g}$ mode of iron ($\omega_{Fe}$) in the samples SmFeAsO$_{1-x}$F$_x$ (x=0, 0.15) with natural abundance oxygen $^nO$ and $^{18}O$, and in the samples SmFeAsO$_{1-x}$F$_x$ (x=0, 0.15) and Ba$_{1-x}$K$_x$Fe$_2$As$_2$ (x=0, 0.4) with $^{54}$Fe and $^{56}$Fe, respectively.** $^{18}O$ and $^{54}$Fe isotopic contents are deduced on the Raman shift being proportional to $1-(16/M')^{1/2}$ for the substitution of $^{18}O$ for $^nO$ and proportional to $1-(55.93/M')^{1/2}$ for the substitution of $^{54}$Fe for $^{56}$Fe, respectively. In the calculation the molar mass for isotope exchange used is 53.93 g for $^{54}$Fe and 55.93 g for $^{56}$Fe, and 16 g for $^nO$ and 18 g for $^{18}O$.

| $\omega_O$ (cm$^{-1}$) | $^nO$ | $^{18}O$ | $\triangle\omega_O$ | $^{18}O\%$ |
|---|---|---|---|---|
| SmFeAsO | 418.6 | 401.2 | -17.4 | 71% |
| SmFeAsO$_{0.85}$F$_{0.15}$ | 425.5 | 406.5 | -19 | 77% |
| $\omega_{Fe}$ (cm$^{-1}$) | $^{56}Fe$ | $^{54}Fe$ | $\triangle\omega_{Fe}$ | $^{54}Fe\%$ |
| SmFeAsO | 208.0 | 211.5 | 3.5 | 92% |
| SmFeAsO$_{0.85}$F$_{0.15}$ | 210.6 | 214.4 | 3.8 | 98% |
| BaFe$_2$As$_2$ | 207.8 | 211.3 | 3.5 | 92% |
| Ba$_{0.6}$K$_{0.4}$Fe$_2$As$_2$ | 204.8 | 208.2 | 3.4 | 91% |

Table 2: **Spin-density-wave and superconducting transition temperature ($T_{SDW}$ and $T_C$), and their shifts ($\triangle T_{SDW}$ and $\triangle T_C$) and isotope exponents ($\alpha_{SDW}$ and $\alpha_C$) for the samples of SmFeAsO$_{1-x}$F$_x$ with natural abundance oxygen $^nO$ and $^{18}O$, and $^{56}Fe$ and $^{54}Fe$, and for the samples Ba$_{1-x}$K$_x$Fe$_2$As$_2$ with $^{54}Fe$ and $^{56}Fe$, respectively.** Isotope exponent is deduced by $\alpha=-d lnT/d lnM$, where M is the atomic mass. To guarantee that one measures the intrinsic isotope effect, isotope back-exchange is usually used: the transition should come back to that before isotope exchange. In our case there is no way to carry out the isotope back-exchange experiment. To get intrinsic isotope effect, we prepared several batches of samples to check the results reproducible from sample to sample. Although $T_C$ slightly varies from batch to batch for SmFeAsO$_{0.85}$F$_{0.15}$, the shift of $T_C$ caused by isotope exchange is nearly the same for all batches. $T_C$ of the sample Ba$_{0.6}$K$_{0.4}$Fe$_2$As$_2$ is very stable and nearly independent of batches. In the calculation the molar masses used are the same as used in Table 1 except for $^{18}O$ which is reduced to 17.42 g and 17.54 g because of only 71% and 77% exchange. The error bar for the temperature determination is equal to the step temperature in the temperature sweep that is 0.02 K for $T_C$ and 0.05 K for $T_{SDW}$ in the resistivity measurements, respectively.

| $T_{SDW}$ | $^nO$ | $^{18}O$ | $\Delta T_{SDW}$ | $\alpha_{SDW}$ |
|---|---|---|---|---|
| SmFeAsO | 130.00(5) | 130.80(5) | 0.80(10) | -0.07(1) |
| | 130.10(5) | 130.60(5) | 0.50(10) | -0.04(1) |
| | 130.10(5) | 130.60(5) | 0.50(10) | -0.04(1) |
| Average | | | | -0.05(1) |
| $T_{SDW}$ | $^{56}Fe$ | $^{54}Fe$ | $\Delta T_{SDW}$ | $\alpha_{SDW}$ |
| SmFeAsO | 130.00(5) | 131.90(5) | 1.90(10) | 0.41(2) |
| | 130.10(5) | 131.80(5) | 1.70(10) | 0.37(2) |
| Average | | | | 0.39(2) |
| BaFe$_2$As$_2$ | 137.00(5) | 138.90(5) | 1.90(10) | 0.39(2) |
| | 137.10(5) | 138.95(5) | 1.85(10) | 0.38(2) |
| | 137.10(5) | 138.70(5) | 1.60(10) | 0.33(2) |
| Average | | | | 0.36(2) |
| $T_C$ | $^nO$ | $^{18}O$ | $\Delta T_C$ | $\alpha_C$ |
| SmFeAsO$_{0.85}$F$_{0.15}$ | 39.80(2) | 40.10(2) | 0.30(4) | -0.08(1) |
| | 41.70(2) | 41.85(2) | 0.15(4) | -0.04(1) |
| | 40.50(2) | 40.85(2) | 0.35(4) | -0.09(1) |
| | 41.50(2) | 41.70(2) | 0.20(4) | -0.05(1) |
| Average | | | | -0.06(1) |
| $T_C$ | $^{56}Fe$ | $^{54}Fe$ | $\Delta T_C$ | $\alpha_C$ |
| SmFeAsO$_{0.85}$F$_{0.15}$ | 41.40(2) | 41.90(2) | 0.50(4) | 0.34(3) |
| | 40.90(2) | 41.40(2) | 0.50(4) | 0.34(3) |
| Average | | | | 0.34(3) |
| Ba$_{0.6}$K$_{0.4}$Fe$_2$As$_2$ | 37.30(2) | 37.80(2) | 0.50(4) | 0.38(3) |
| | 37.30(2) | 37.82(2) | 0.52(4) | 0.39(3) |
| | 37.40(2) | 37.86(2) | 0.46(4) | 0.35(3) |
| Average | | | | 0.37(3) |

**Figure legends:**

Figure 1: **Raman spectra at room temperature for the samples SmFeAsO$_{1-x}$F$_x$ (x=0 and 0.15) and Ba$_{1-x}$K$_x$Fe$_2$As$_2$ (x=0 and 0.4)**. SmFeAsO$_{1-x}$F$_x$ (x=0 and 0.15), (a): with $^n$O and $^{18}$O, (b): with $^{56}$Fe and $^{54}$Fe; (c): Ba$_{1-x}$K$_x$Fe$_2$As$_2$ (x=0 and 0.4) with $^{56}$Fe and $^{54}$Fe. The mode around 420 cm$^{-1}$ is ascribed to the E$_g$ mode of oxygen and the mode around 210 cm$^{-1}$ is assigned to the B$_{1g}$ mode of Fe[18].

Figure 2: **Temperature dependence of resistivity (ρ) and its derivative (dρ/dT) for the samples SmFeAsO$_{1-x}$F$_x$ with $^n$O and $^{18}$O.** The peak in the derivative is considered as the transition temperature. The peak temperature in dρ/dT corresponds to mid-transition temperature for superconducting transition. An anomalous peak at about 147 K in resistivity is associated with the structural transition for the x=0 sample, and the peak temperature in dρ/dT is very close to the SDW ordering temperature observed by neutron scattering[19].

Figure 3: **Temperature dependence of resistivity (ρ) and its derivative (dρ/dT) for the samples SmFeAsO$_{1-x}$F$_x$ and Ba$_{1-x}$K$_x$Fe$_2$As$_2$ isotopically substituted with $^{56}$Fe and $^{54}$Fe.** (a): SmFeAsO$_{1-x}$F$_x$ with x=0; (b): SmFeAsO$_{1-x}$F$_x$ with x=0.15; (c): Ba$_{1-x}$K$_x$Fe$_2$As$_2$ with x=0; (d): Ba$_{1-x}$K$_x$Fe$_2$As$_2$ with x=0.4. In order to accurately determine the transition temperature, we took the derivative of the resistivity, and the peak in the derivative is considered as the transition temperature. The peak temperature in dρ/dT for the SDW is very close to the ordering temperature

observed by neutron scattering[19]. The $\alpha_C$ and $\alpha_{SDW}$ are deduced to be 0.34(3) and 0.37(2) for the samples SmFeAsO$_{1-x}$F$_x$ with x=0.15 and 0, respectively. For a typical sample of BaFe$_2$As$_2$, the resistivity anomaly around 145 K arises from the structural and SDW transition[20]. For a typical superconducting sample of Ba$_{1-x}$K$_x$Fe$_2$As$_2$ with x=0.4, the onset transition occurs around 39 K. The $\alpha_C$ and $\alpha_{SDW}$ are found to be 0.38(3) and 0.39(2) for the samples Ba$_{1-x}$K$_x$Fe$_2$As$_2$ with x=0.4 and 0, respectively. To confirm the results from resistivity measurements, the susceptibility for the same samples is measured (see Supplementary Figure S2). The shift of $T_C$ and $T_{SDW}$ determined from susceptibility measurements due to iron isotope exchange is consistent with that obtained from the resistivity measurements. It should be pointed out that a remarkable difference in resistivity is observed below SDW transition between the x=0 samples with $^{56}$Fe and $^{54}$Fe for the SmFeAsO$_{1-x}$F$_x$ and Ba$_{1-x}$K$_x$Fe$_2$As$_2$ systems. As shown in the inset of Fig.3a and 3c, the resistivity decreases by about 28% for SmFeAsO sample and by about 35% for BaFe$_2$As$_2$ sample at 10 K due to iron isotope exchange. It suggests that iron isotope exchange has strong effect on spin-density-wave state.

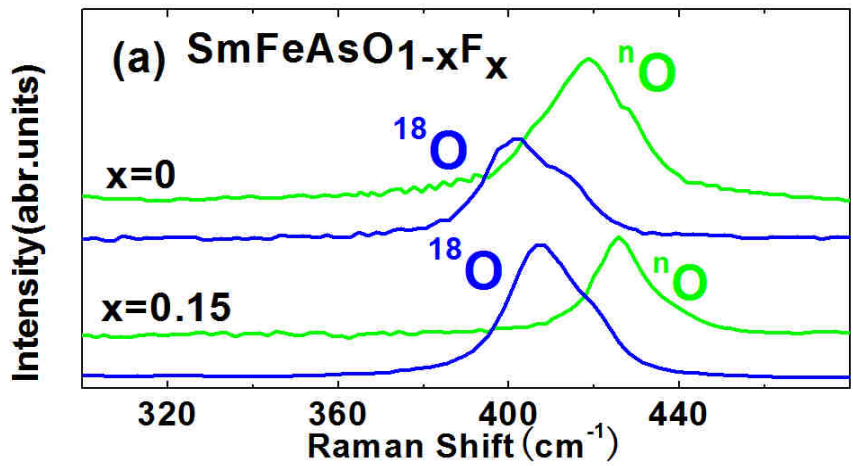
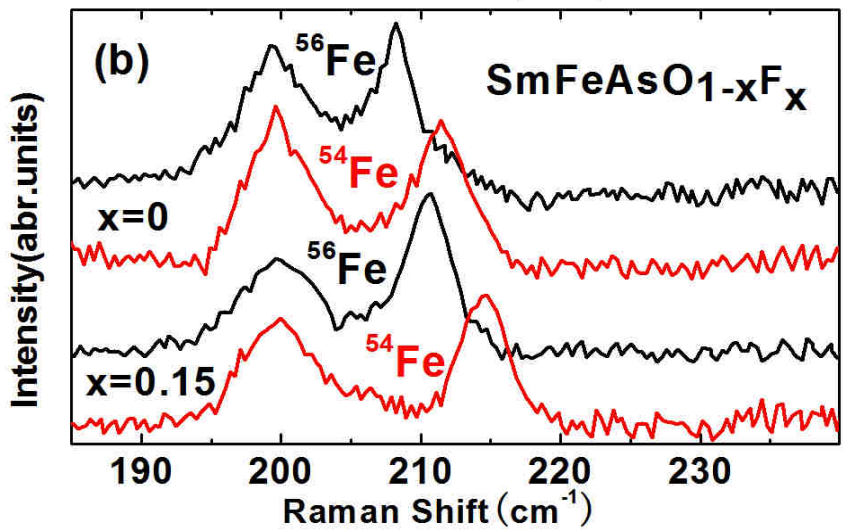
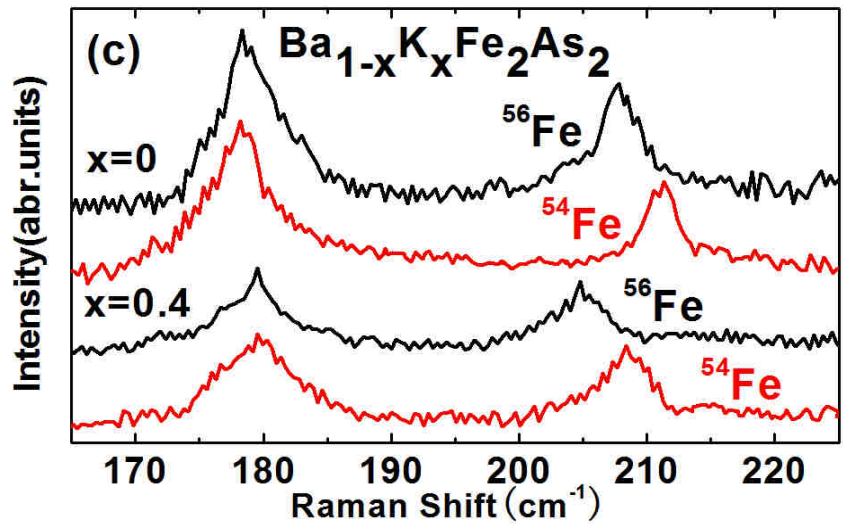

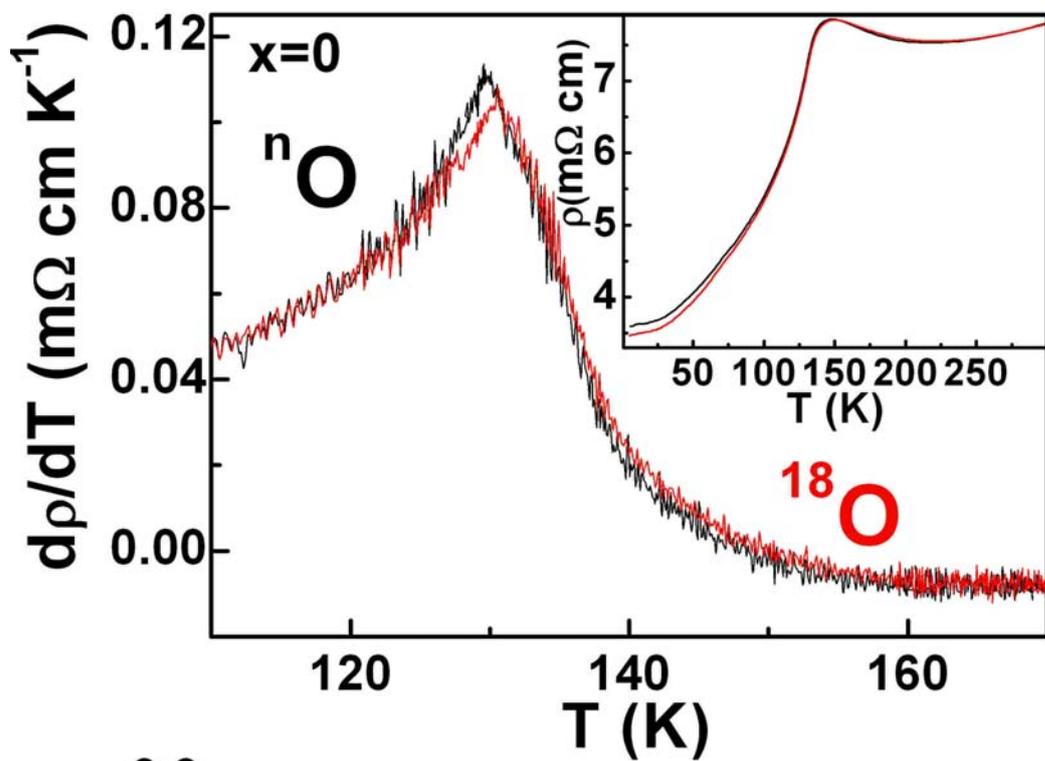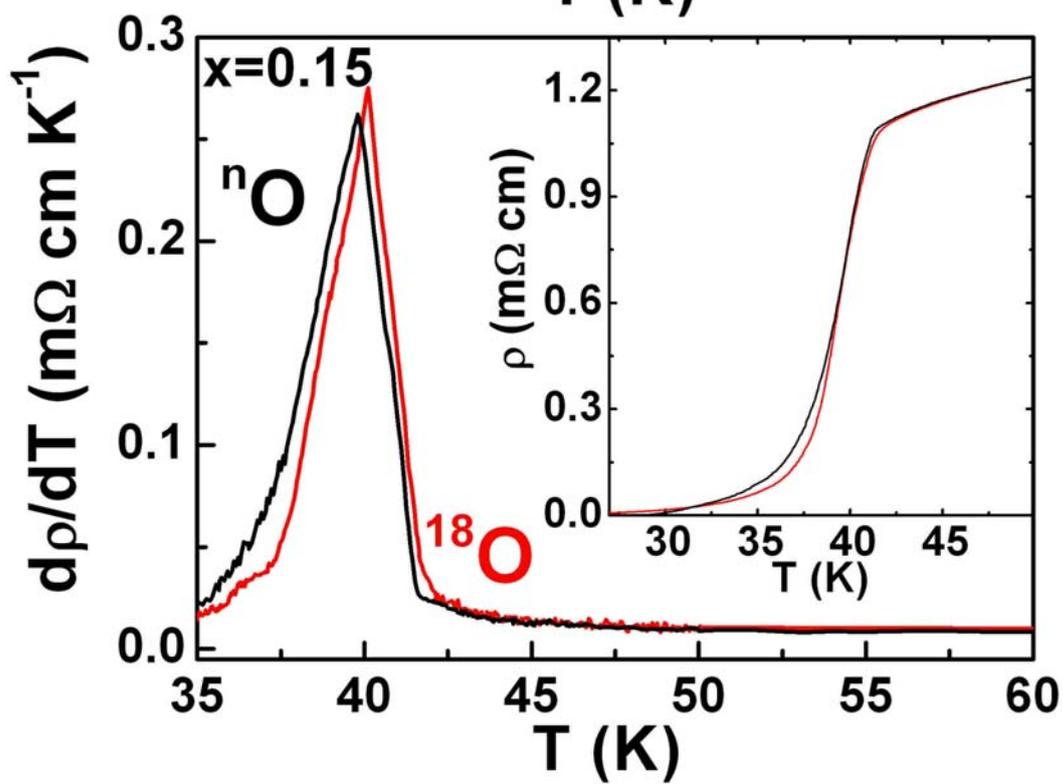

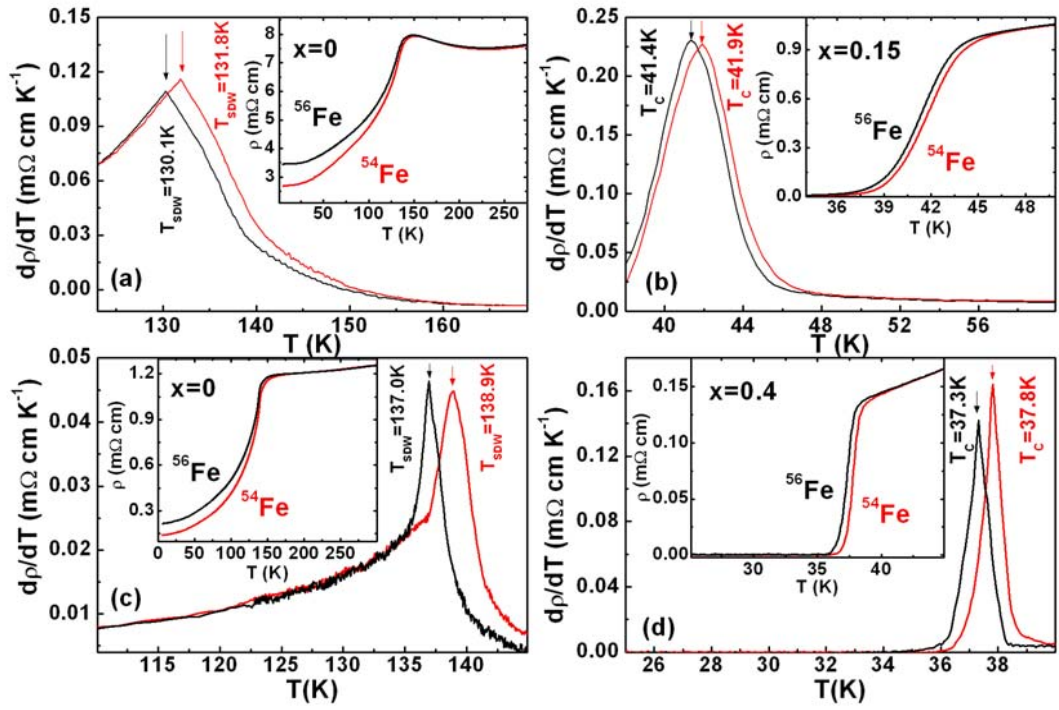

# Supplementary Information

The isotopically substituted polycrystalline samples with nominal compositions SmFeAsO$_{1-x}$F$_x$ (x=0, 0.15) were synthesized by conventional solid state reaction using SmAs, SmF$_3$, As, Fe and Fe$_2^n$O$_3$/Fe$_2^{18}$O$_3$ as starting materials for oxygen isotope exchange, while using SmAs, SmF$_3$, As, $^{54}$Fe/$^{56}$Fe and $^{54}$Fe$_2$O$_3$/$^{56}$Fe$_2$O$_3$ as starting materials for iron isotope exchange. Fe$_2^n$O$_3$ and Fe$_2^{18}$O$_3$ were obtained by sintering Fe powder in natural abundance oxygen $^n$O$_2$ and $^{18}$O$_2$ with an 80% enrichment, respectively. The $^{54}$Fe$_2$O$_3$ and $^{56}$Fe$_2$O$_3$ were prepared by sintering $^{54}$Fe and $^{56}$Fe powder in flowing natural abundance oxygen atmosphere, respectively. The iron isotope polycrystalline samples of Ba$_{1-x}$K$_x$Fe$_2$As$_2$ (x=0, 0.4) were synthesized using BaAs, KAs, and $^{56}$Fe$_2$As and $^{54}$Fe$_2$As as starting materials, respectively. $^{56}$Fe$_2$As and $^{54}$Fe$_2$As were synthesized by reacting $^{56}$Fe and $^{54}$Fe powder with As powder in evacuated quartz tubes at 650 $^o$C for 24 h, respectively. To keep the same condition, the isotope exchange samples with the same composition were sealed in the same evacuated quartz tube for annealing. To mitigate the difference of F content in isotope exchange samples SmFeAsO$_{0.85}$F$_{0.15}$, SmAs, SmF$_3$ and As were first mixed according to stoichiometric ratio and grounded, then mixture of SmAs, SmF$_3$ and As was equally separated into two parts. Finally, $^{54}$Fe/$^{56}$Fe and $^{54}$Fe$_2$O$_3$/$^{56}$Fe$_2$O$_3$ as starting materials were weighed and put into the separated mixture of SmAs, SmF$_3$ and As for iron isotope exchange. It guarantees the same F content for isotope exchange in the beginning of sample process. They were loaded into the same quartz tube for annealing.

Natural abundance iron was used for the oxygen isotope exchange experiments and obtained from

Alfa Aesar, while the $^{56}$Fe and $^{54}$Fe enriched isotopes from ISOFLEX (San Francisco, USA) were used for the iron isotope experiments. The purity of the natural abundance iron $^{n}$Fe is 99.9%, while the purity of the $^{56}$Fe (enrichment: 97%+) is 99.78%, and the purity of the $^{54}$Fe (enrichment: 96%+) is 99.86%. Natural abundance oxygen was used for the iron isotope exchange experiments for the system SmFeAsO$_{1-x}$F$_x$. The purity of the natural abundance oxygen $^{n}$O is 99.99%. The purity of the $^{18}$O (enrichment: 80%) is 99.99%. Raman spectra were obtained on a LABRAM-HR Confocal Laser MicroRaman Spectrometer using the 514.5 nm line from an argon-ion laser with in-plane light polarization. X-ray diffraction (XRD) was performed by MAC MXPAHF X-Ray diffractometer with graphite monochromated CuK$_{\alpha 1}$ radiation ($\lambda$=0.15406 nm) at room temperature. The superconducting transition temperatures and the SDW transition temperatures were determined by standard four-probe resistance and susceptibility measurements. The resistance was measured by an AC resistance bridge (LR-700, Linear Research). Magnetic susceptibility measurements were performed with a superconducting quantum interference device magnetometer (Quantum Design MPMS-7).

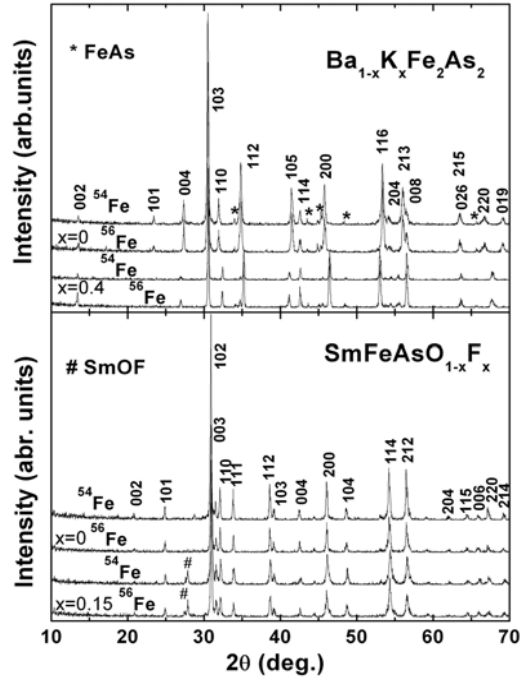

Figure S1: **X-ray diffraction patterns at room temperature for the samples SmFeAsO$_{1-x}$F$_x$ and Ba$_{1-x}$K$_x$Ba$_2$As$_2$ with iron isotope exchange.** As shown in the Figure, XRD patterns look like the same for isotope exchange. The XRD patterns are fitted by Rietveld analysis. The obtained lattice constants are: a=0.3961(1) nm, c=1.3015(4) nm for BaFe$_2$As$_2$ with $^{54}$Fe, a=0.3962(1) nm, c=1.3017(4) nm for BaFe$_2$As$_2$ with $^{56}$Fe; a=0.3918(2) nm, c=1.3289(7) nm for Ba$_{0.6}$K$_{0.4}$Fe$_2$As$_2$ with $^{54}$Fe, a=0.3917(2) nm, c=1.3288(7) nm for Ba$_{0.6}$K$_{0.4}$Fe$_2$As$_2$ with $^{56}$Fe; a=0.39415(6) nm, c=0.8502(3) nm for SmFeAsO with $^{54}$Fe, a=0.39410(6) nm, c=0.8501(3) nm for SmFeAsO with $^{56}$Fe; a=0.39320(5) nm, c=0.8490(2) nm for SmFeAsO$_{0.85}$F$_{0.15}$ with $^{54}$Fe, a=0.39326(5) nm, c=0.8491(2) nm for SmFeAs O$_{0.85}$F$_{0.15}$ with $^{56}$Fe, respectively. The lattice constants are the same within experimental error.

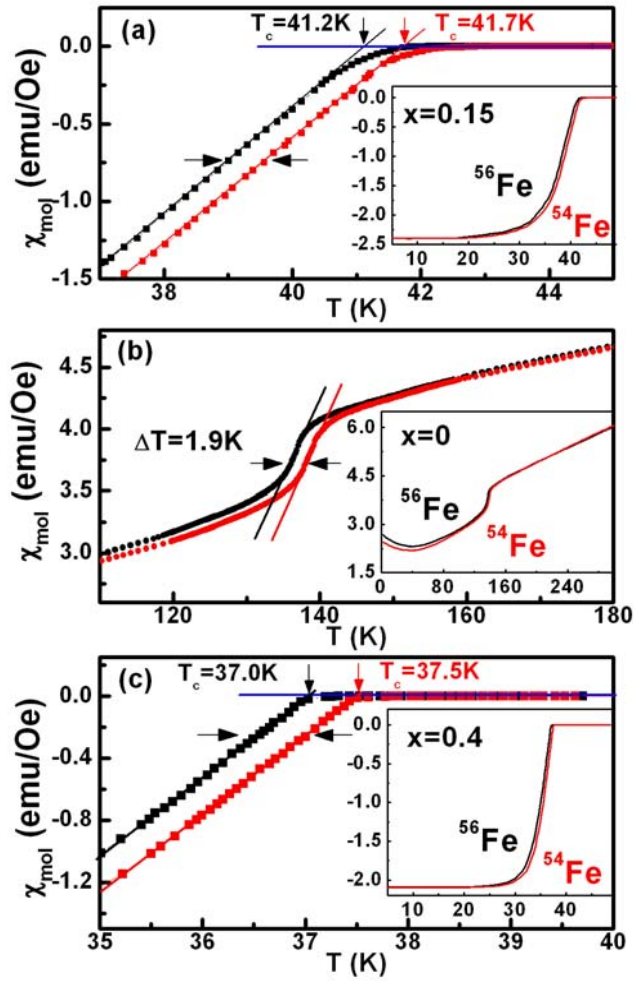

Figure S2: **Temperature dependence of susceptibility for the typical samples of SmFeAsO$_{0.85}$F$_{0.15}$ and Ba$_{1-x}$K$_x$Fe$_2$As$_2$ isotopically substituted with $^{56}$Fe and $^{54}$Fe.** (a): SmFeAsO$_{0.85}$F$_{0.15}$; (b): Ba$_{1-x}$K$_x$Fe$_2$As$_2$ with x=0; (c): Ba$_{1-x}$K$_x$Fe$_2$As$_2$ with x=0.4. To confirm the results from resistivity measurements, the susceptibility of the same samples used in Figure 3 is measured. The shift of T$_C$ and T$_{SDW}$ determined from susceptibility measurements is consistent with that obtained from the resistivity measurements as shown in Figure 3. The susceptibility of superconducting samples SmFeAsO$_{0.85}$F$_{0.15}$ and Ba$_{0.6}$K$_{0.4}$Fe$_2$As$_2$ is measured in the zero-field cooling process under the magnetic field of 10 Oe. The susceptibility of parent compound BaFe$_2$As$_2$ is measured in the field-cooled process under the magnetic field of 50000 Oe.

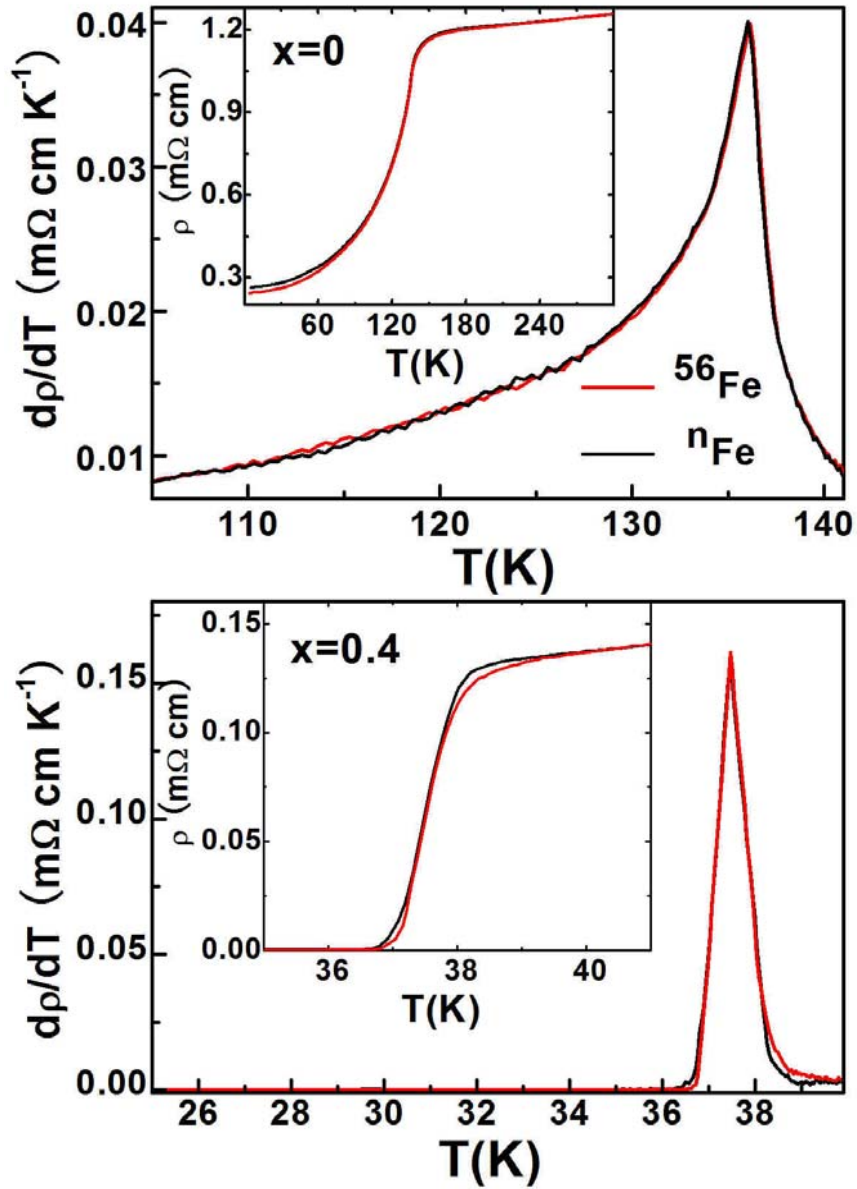

Figure S3: **Temperature dependence of resistivity (ρ) and its derivative (dρ/dT) for the samples $Ba_{1-x}K_xFe_2As_2$ (x=0 and 0.4) with nature abundance iron $^n$Fe and $^{56}$Fe .** The natural abundance iron $^n$Fe with a 91.8% enrichment of $^{56}$Fe and purity of 99.9%; the enrich iron $^{56}$Fe with a 97%+ enrichment and purity of 99.78%. It clearly shows that no apparent change in $T_{SDW}$ and $T_C$ is observed in the samples with nature abundance iron $^n$Fe and enriched iron $^{56}$Fe.

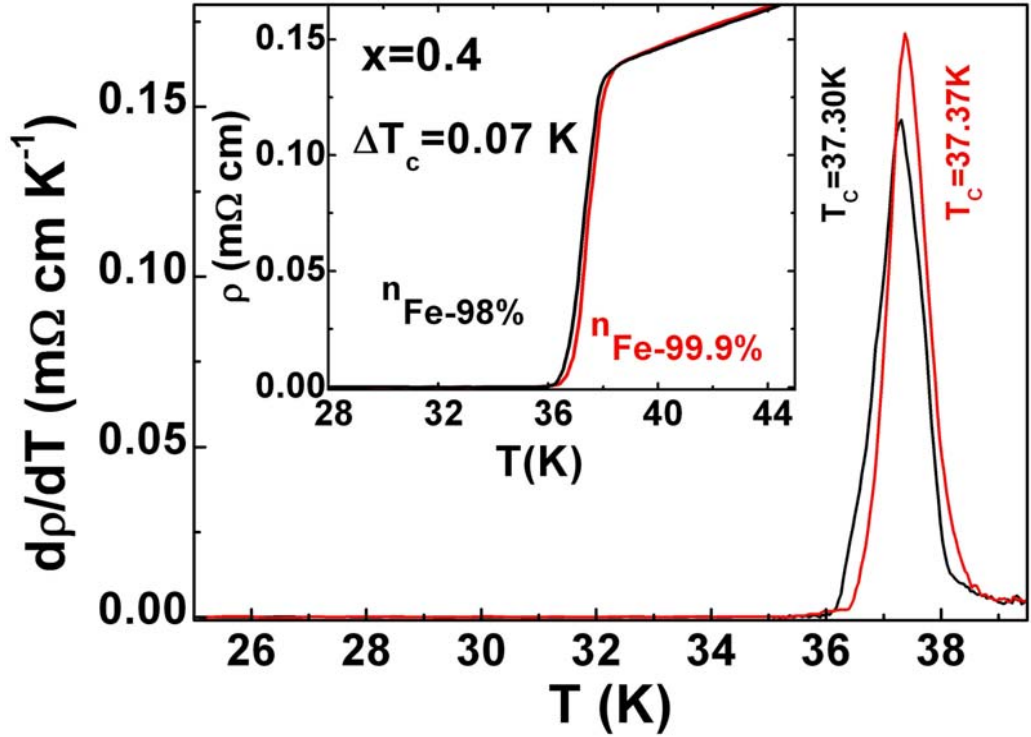

Figure S4: **Temperature dependence of resistivity (ρ) and its derivative (dρ/dT) for the samples $Ba_{0.6}K_{0.4}Fe_2As_2$ with nature abundance irons $^nFe$ with purity of 98％ and 99.9%, respectively.** To check the effect of impurity on $T_C$, the samples with different purity of iron $^nFe$ are studied. It clearly shows that the shift of $T_C$ for the two samples with different purity of iron is about 0.07 K. It indicates that the effect of impurity on $T_C$ is very little. In addition, the purity of isotope iron used in our work is 99.78% for $^{56}Fe$ and 99.86% for $^{54}Fe$, so that the change of Tc from impurity effect should be much less than 0.07 K due to smaller purity difference. Therefore, the effect of impurity on $T_C$ can be ignored, and it further indicates that the isotope effect reported here is intrinsic.